\documentclass[twocolumn,floatfix,showpacs,amsmath,amssymb,prl]{revtex4}
\usepackage{times}
\usepackage{epsf}
\usepackage{epsfig}
\usepackage{graphicx}
\usepackage{dcolumn}
\usepackage{bm}

\def\s0#1#2{\mbox{\small{$ \frac{#1}{#2} $}}}
\def\0#1#2{\frac{#1}{#2}}
\def\eq#1{(\ref{#1})}
\def\Eq#1{Eq.~(\ref{#1})}

\begin{document}

\title{Functional renormalisation group approach to far-from-equilibrium quantum field dynamics}
\author{Thomas Gasenzer}
\thanks{email:T.Gasenzer@thphys.uni-heidelberg.de}
\author{Jan M. Pawlowski}
\thanks{email:J.Pawlowski@thphys.uni-heidelberg.de}
\affiliation{Institut f\"ur Theoretische Physik, Universit\"at
  Heidelberg, Philosophenweg 16, 69120 Heidelberg, Germany}

\begin{abstract}
\noindent
Dynamic equations for quantum fields far from equilibrium are derived
by use of functional renormalisation group techniques.  The obtained
equations are non-perturbative and lead substantially beyond
mean-field and quantum Boltzmann type approximations.  The approach is
based on a regularised version of the generating functional for
correlation functions where times greater than a chosen cutoff time
are suppressed.  As a central result, a time evolution equation for
the non-equilibrium effective action is derived, and the
time evolution of the Green functions is computed within a vertex
expansion. It is shown that this agrees with the dynamics derived from
the $1/\cal N$-expansion of the two-particle irreducible effective
action.

\end{abstract}
\pacs{
03.75.Kk, 
05.10.Cc, 
05.30.-d, 
05.70.Ln, 
11.15.Pg  
\hfill HD--THEP--07--31}

\maketitle
\textit{Introduction}.
\label{sec:intro}
Far-from-equilibrium quantum field dynamics is one of the most
challenging issues both in experimental and theoretical physics to
date.  Experiments exhibiting quantum statistical effects in the time
evolution of many-body systems are extremely demanding.  In
particular, the preparation of ultracold atomic Bose and Fermi gases
in various trapping environments allows to precisely study quantum
many-body dynamics of strongly correlated systems, see, e.g.,
Refs.~\cite{Greiner2002a}.  In recent years, the field has attracted
researchers from a variety of disciplines, ranging from
condensed-matter to high-energy particle physics and cosmology.
Nonequilibrium field theory to date is dominated by semi-clas\-sical
mean-field approaches which are in general only valid for weak
interactions or large occupation numbers.  For strongly correlated
quantum systems methods are available predominantly for systems in one
spatial dimension and include the Density Matrix Renormalisation Group
methods, see e.g.~\cite{Vidal2004a}, as well as techniques for exactly
solvable models \cite{Korepin1997a}.  Non-perturbative approximations
of the two-particle irreducible (2PI) effective action
\cite{Cornwall1974a} have been intensively studied and applied to
nonequilibrium dynamics
\cite{Berges2002a,Aarts:2002dj,Mihaila:2003mh,Berges:2002wr,%
  Gasenzer:2005ze,Aarts:2006cv,Berges2007a} and are applicable also in
more than one dimension.  In field theory, they also provide a way to
study strongly correlated fermions beyond mean-field and Boltzmann
approximations \cite{Berges:2002wr}.  Like these, the results
presented here are expected to be of high relevance, e.g., for the
description of ultracold degenerate Fermi gases close to the BEC-BCS
crossover \cite{Regal2004b}.

In this paper we derive dynamic equations for quantum fields far from
equilibrium. As a central result, we derive a new time evolution
equation for the non-equilibrium effective action by use of functional
renormalisation group (RG) techniques,
cf.~Refs.~\cite{Wetterich:1992yh,Bagnuls:2000ae,Litim:1998nf,
  Pawlowski:2005xe}, as well as \cite{Canet:2003yu} for
non-equilibrium applications. This exact and closed evolution equation
allows for non-perturbative approximations that lead substantially
beyond mean-field and quantum Boltzmann type approaches: it can be
rewritten, in a closed form, as a hierarchy of dynamical equations for
Green functions, and this hierarchy admits truncations that neither
explicitly nor implicitly rely on small bare couplings or
close-to-equilibrium evolutions.  Moreover, the hierarchy of equations
is technically very close to 2PI and Dyson-Schwinger type evolutions,
see \cite{Pawlowski:2005xe}, as well as to evolution equations for the
effective action as derived in \cite{Wetterich:1996ap}. This has the
great benefit that results from either method can be used as an input
as well as for reliability checks of the respective truncation
schemes. In turn, in particular the different resummation schemes
invoked in these approaches, as well as the differing dependences on
Green functions, allow for a systematic analysis of the mechanisms of
non-equilibrium physics.  Here we apply the above setting within a
non-perturbative vertex expansion scheme as well as an $s$-channel
approximation. This truncation turns out to correspond to an expansion
in inverse powers of the number of field components $\cal N$. The
dynamic equations derived agree to next-to-leading order with those
obtained from a $1/\cal N$ expansion of the 2PI effective action.

\textit{Functional renormalisation group approach}.
\label{sec:RGapproach}
For a given initial-state density matrix $\rho_D(t_0)$, the
renormalised finite quantum generating functional for time-dependent
$n$-point correlation functions,
\begin{align}\label{eq:definingZneq}
  Z[J;\rho_D] &= {\rm Tr}\Big[\rho_D(t_0)\, {\cal T}_{\cal C}
  \exp\Big\{i \int_{x,{\cal C}}\! J_a(x) \Phi_a(x) \Big\}\Big],
\end{align}
(summation over double indices is implied) carries all the information
of the quantum many-body evolution at times greater than the initial
time $t_0$.  The Heisenberg field operators $\Phi_a(x)$,
$a=1,...,{\cal N}$, are assumed to obey equal-time bosonic commutation
relations.  In Eq.~(\ref{eq:definingZneq}), ${\cal T}_{\cal C}$
denotes time-ordering along the Schwinger-Keldysh closed time path
(CTP) $\cal C$ leading from $t_0$ along the real-time axis to infinity
and back to $t_0$, with $\int_{x,{\cal C}} \equiv\int_{\cal C}
{\mathrm d} x_0 \int {\mathrm d}^d x$.  All connected Greens functions
will be time-ordered along $\cal C$.

\begin{figure}[htb]
\begin{center}
\includegraphics[width=0.47\textwidth]{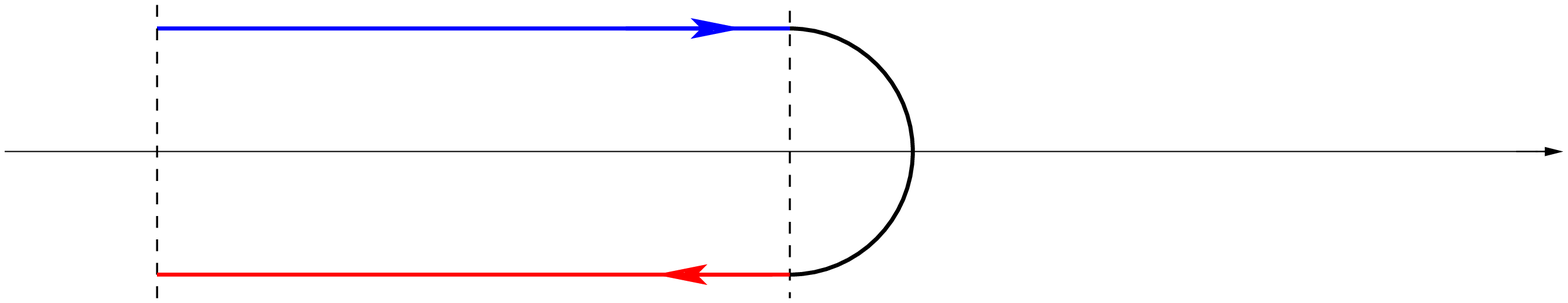}
\newline
\begin{picture}(0,0)(0,0)
\put(116,40){$t$}
\put(-112,40){$t_0$}
\put(-10,40){$\tau$}
\end{picture}
\end{center}
\vspace*{-3ex} \caption{(color online) The closed time path ${\cal
    C}(\tau)$ terminating at the time given by the parameter $\tau$.
  At later times, field fluctuations summed over in the generating
  functional do not contribute to Green functions the maximum time
  argument of which is $\tau$.}
\label{fig:CTPtau}
\end{figure}
The key idea of our approach to dynamics is to first consider the
generating functional for Green functions where all times are smaller
than a maximum time $\tau$. This implies a time path ${\cal C}(\tau)$
which is closed at $t=\tau$, see Fig.~\ref{fig:CTPtau}, and we are led
to the generating functional $Z_\tau=Z_{{\cal C}(\tau)}$ with the
source term
\begin{equation}
 {\cal T}_{{\cal C}(\tau)}
  \exp\Big\{i \int_{x,{\cal C}(\tau)}\! J_a(x) \Phi_a(x) \Big\}\,. 
\end{equation}
At $\tau =t_0$, this results in a trivial $Z_{t_0}$ where all
information is stored in the initial density matrix $\rho_D(t_0)$.
From this initial condition $Z_\tau$ can be computed by means of the
time evolution $\partial_\tau Z_\tau$ for all times $\tau>t_0$.  

\begin{figure}[htb]
\begin{center}
\includegraphics[width=0.3\textwidth]{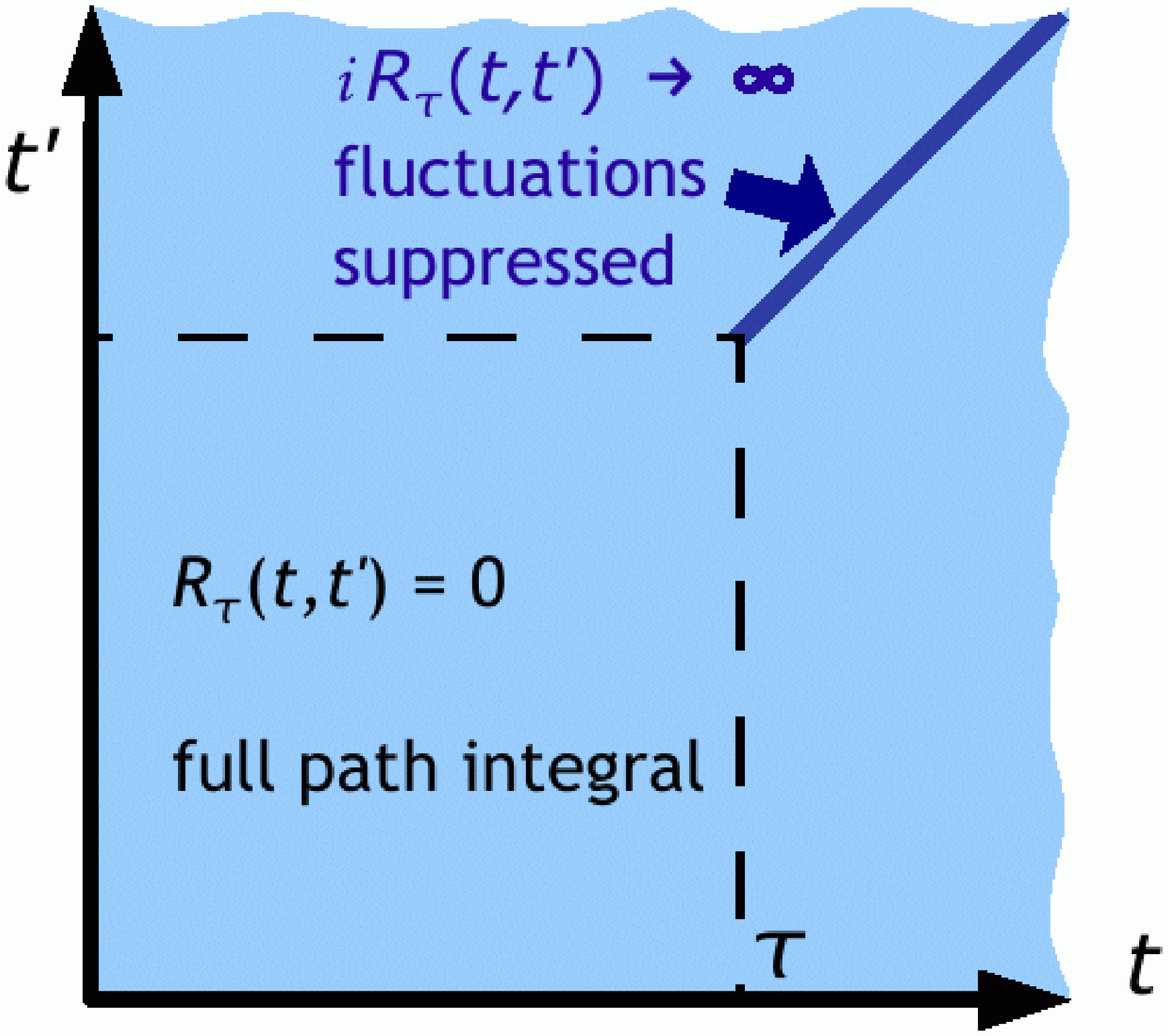}
\end{center}
\vspace*{-3ex} 
\caption{(Color online) The cutoff function $R_{\tau,ab}(x,y)$ in the time
plane $\{x_0,y_0\}=\{t,t'\}$, $t,t'\ge t_0$. The function vanishes everywhere 
except for $t=t'>\tau$ where it tends to infinity and therefore 
implies a suppression of all 
fluctuations in the generating functional at times greater than $\tau$.}
\label{fig:Rchoice}
\end{figure}

We will derive this evolution by using functional RG ideas.  To that
end we note that $Z_\tau$ can be defined in terms of the full
generating functional $Z_\infty$ in \eq{eq:definingZneq} by
suppressing the propagation for times greater than $\tau$. This
suppression is achieved by
\begin{align}
\label{eq:defZtau}
  Z_\tau &= \exp\Big\{-\frac{i}{2} \int_{x y,{\cal C}}\!
  \frac{\delta}{\delta J_a(x)}R_{\tau,ab}(x,y) \frac{\delta}{\delta
  J_b(y)}\Big\}Z,
\end{align}
where the function $R_\tau$ is chosen such that it suppresses the
fields, i.e., $\delta/\delta J_a$, for all times $t> \tau$. This
requirement does not fix $R_\tau$ in a unique way, and a simple choice 
is provided by 
\begin{equation}
\label{eq:Rchoice}
  -i R_{\tau,ab}(x,y)
  =\left\{\begin{array}{lcl} 
          \infty & \quad & \mathrm{for}\ x_0=y_0>\tau, 
	                   \mathbf{x}=\mathbf{y}, a=b \\[2ex] 
          0  	 & 		 & \mathrm{otherwise} 
          \end{array} \right.,
\end{equation}
see Fig.~\ref{fig:Rchoice}. We note that for $\tau\to\infty$, we
recover the full generating functional, $Z_\infty\equiv Z$, while for
$\tau=t_0$, any evolution to times $t>t_0$ is suppressed.  In the
latter case all $n$-point functions derived from $Z_\tau$ are defined
at $t_0$ only and reduce to the free classical ones, or any other
physical boundary condition.

\noindent We emphasise that the cutoff $R_\tau$ in (\ref{eq:defZtau})
suppresses any time evolution at times greater than $\tau$.
Correlation functions derived from $Z_\tau$ vanish as soon as at least
one of their time arguments is larger than $\tau$.  Hence, the
regularised generating functional (\ref{eq:defZtau}) is equivalent to
a generating functional with a closed time path ${\cal C}(\tau)$
leading from $t_0$ to $\tau$ and back to $t_0$.  Note that the CTP
automatically arranges for the normalisation of $Z_\tau$. We conclude
that, by construction, the sharp cutoff (\ref{eq:Rchoice}) is a
physical one in the sense that it corresponds to integrating out all
fluctuations being relevant for the evolution up to a particular time.

The restriction of the CTP to times $t_0\le t\le\tau$ implies that the
differential equation for $Z_\tau$ describing the flow of the
generating functional, and therefore that of the correlation
functions, encodes the full time evolution of the system. Analogously,
the time evolution of connected correlation functions is derived from
that of the Schwinger functional $W_\tau=-i \ln Z_\tau$. It is more
convenient, however, to work with the effective action,
\begin{align}
  \label{eq:effAction}
  \Gamma_\tau[\phi;R_\tau]
  &= W_\tau[J;\rho_D]-\int_{\cal C}J_a\phi_a
      - \frac{1}{2}\int_{\cal C}\phi_a  R_{\tau,ab}\phi_b.
\end{align}
Here, space-time arguments are suppressed, and $\phi_a(x)=\delta
W_\tau/\delta J_a(x)|_{J\equiv0}$ is the classical field expectation
value. From Eqs.~\eq{eq:defZtau} and \eq{eq:effAction} we derive the
Functional RG or flow equation for the $\tau$-dependent effective
action,
\begin{align}
  \label{eq:flowGamma}
  \partial_\tau \Gamma_\tau
  &= \frac{i}{2} \int_{{\cal C}}\!
  \left[\0{1}{\Gamma^{(2)}_\tau+R_\tau}\right]_{ab}
  \partial_\tau R_{\tau,ab}\, ,
\end{align}
where $\Gamma_\tau^{(n)}=\delta^n \Gamma_\tau/(\delta \phi)^n$.
Again, space-time arguments are suppressed which appear in analogy to
the field indices $a,b$, see e.g.~Ref.~\cite{Pawlowski:2005xe}.
\Eq{eq:flowGamma} represents our central result.  It is analogous to
functional flow equations used extensively with regulators in momentum
and/or frequency space to describe strongly correlated systems near
equilibrium
\cite{Wetterich:1992yh,Bagnuls:2000ae,Litim:1998nf,Pawlowski:2005xe}.
Its homogenous part relates to standard $\tau$-dependent
renormalisation \cite{Pawlowski:2005xe}, and has been studied eg.\ in
\cite{Boyanovsky:1998aa,Ei:1999pk}. We close the derivation of the
evolution equation with some remarks: \Eq{eq:flowGamma} is nothing but
an infinitesimal closed time loop at $\tau$, which is by itself finite
and requires no further renormalisation. However, in particular in
higher dimensions additional regulators in the spatial or momentum
domains may be advantageous for facilitating renormalisation and thus
the practical application of the approach, see
\cite{Wetterich:1992yh,Bagnuls:2000ae,Pawlowski:2005xe}. We also
emphasise that the evolution equation \eq{eq:flowGamma}, even though
close in spirit and construction to standard functional RG equations,
is conceptually different. It entails a physical time evolution as
opposed to integrating out degrees of freedom.  Nevertheless, more
general regulators $R_\tau$ may be advantageous in other cases
\cite{Pawlowski:2005xe}, e.g., when a cutoff is set in the relative
time ($t-t'$-) direction, see Fig.~\ref{fig:Rchoice}. Such a case
would correspond to setting a cutoff along the frequency axis
\cite{Canet:2003yu} which is conceptually closer to the standard RG
approach with cutoffs in the momentum domain. A more detailed
discussion of our approach is deferred to \cite{Gasenzer2008a}.

\begin{figure}[htb]
\begin{center}
\includegraphics[width=0.49\textwidth]{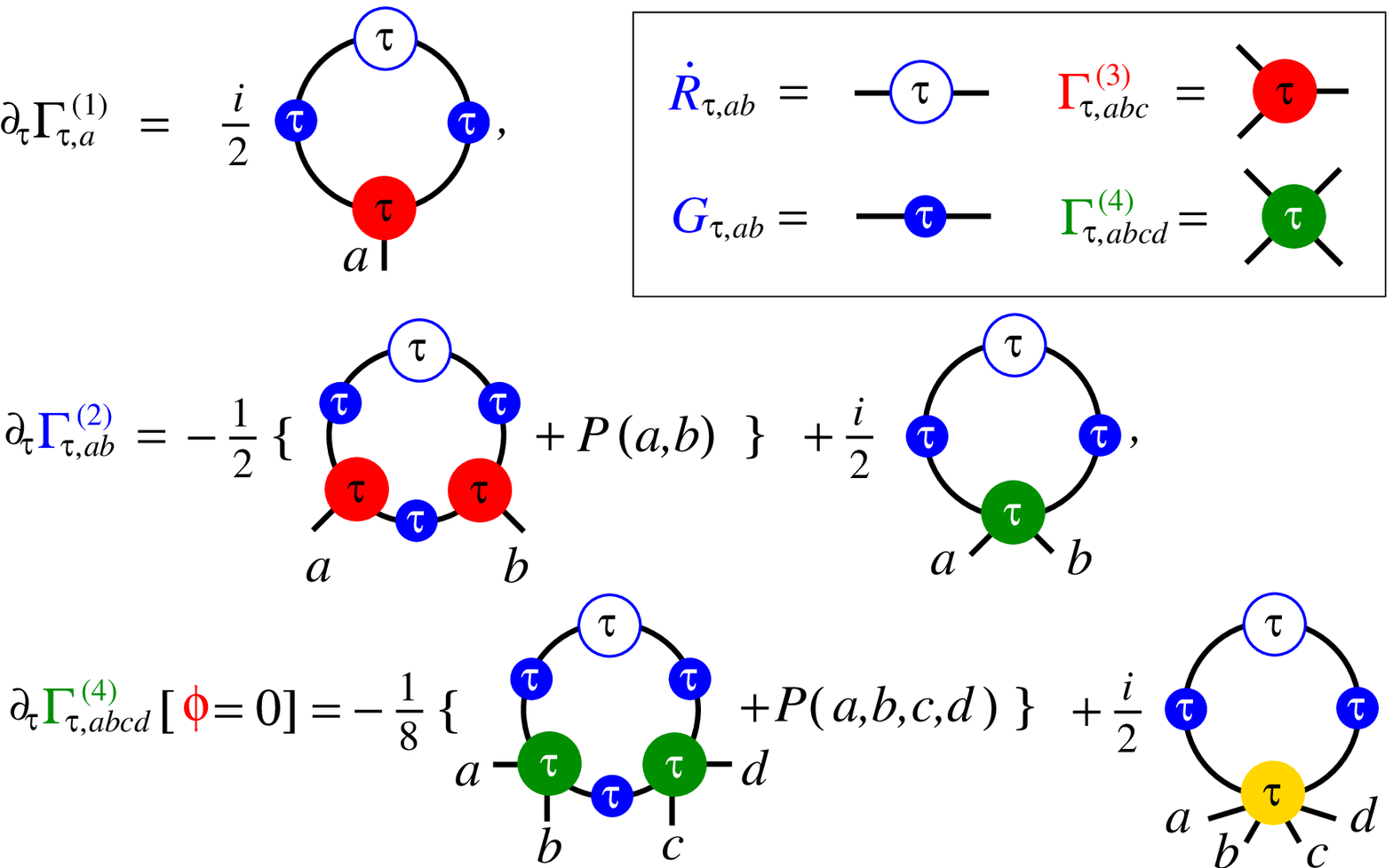}
\end{center}
\vspace*{-3ex} \caption{(color online) Diagrammatic representation of
  the general flow equations for
  $\Gamma_\tau^{(1)}[\phi],\Gamma_\tau^{(2)}[\phi]$, and
  $\Gamma_{\tau,abcd}^{(4)}[\phi=0]$, for a $\phi^4$-theory.  Open
  circles with a $\tau$ denote $\partial_\tau R_{\tau,ab}$, solid
  lines with (blue) filled circles are $\tau$- and, in general,
  $\phi$-dependent two-point functions
  $G_{\tau,ab}=i[\Gamma_\tau^{(2)}+R_\tau]^{-1}_{ab}$.  All other
  filled circles denote proper field-dependent $n$-vertices
  $\Gamma^{(n)}_{\tau,abcd}$, $n=3,4,6$. $P$ implies a sum
  corresponding to all permutations of its arguments. }
\label{fig:FlowEqs}
\end{figure}

\textit{Flow equations for correlation functions}.
\label{sec:FlowEqs}
To obtain a practically solvable set of dynamic equations, we derive 
the flow equation for the proper $n$-point Green function
$\Gamma_{\tau}^{(n)}$ by taking the $n$th field derivative of
Eq.~\eq{eq:flowGamma}.  Fig.~\ref{fig:FlowEqs} shows a diagrammatic
representation of the resulting equations for the $\tau$-dependent
proper two- and four-point functions.

\noindent We would like to emphasise that proper $n$-point Green
functions in general do not vanish for times greater than $\tau$ as
opposed to the correlation functions derived from $Z_\tau$.  However,
in contrast to usual renormalisation-group flows in the momentum
domain, causality prevents the influence of proper Green functions for
times greater than $\tau$ on the dynamics up to time $\tau$.

To be more specific, we consider in the following the special case
of an $\cal N$-component scalar $\phi^4$ theory defined by the
classical action
\begin{align}
  S[\varphi]
  &= \frac{1}{2}\int_{x y} \varphi_a(x) iG_{0,ab}^{-1}(x,y)\varphi_b(y)
    -\frac{g}{4{\cal N}} \int_{x} (\varphi(x)^2)^2,
\label{eq:Sq}
\end{align}
where $\varphi^2=\varphi_a\varphi_a$, and where the inverse free
classical propagator involves one or more time derivatives, e.g.,
$G^{-1}_{0,ab}(x,y)=\delta_{\cal C}(x-y)[-\sigma^2_{ab}\partial_{x_0}
+iH_\mathrm{1B}(x)\delta_{ab}]$ for a non-relativistic trapped Bose
gas, with $\sigma^2$ being the Pauli 2-matrix and
$H_\mathrm{1B}(x)=-\Delta_{\mathbf{x}}^2/2m+V(x)$ the one-body
Hamiltonian.

Our goal is to derive the full time-evolution of
$\Gamma^{(n)}=\Gamma^{(n)}_{\infty}$, in particular, of the connected
two-point function
$G=i[\Gamma^{(2)}]^{-1}=i[\Gamma^{(2)}_{\infty}]^{-1}$ which contains
all information about the normal and anomalous one-body density
matrices (cf., e.g.~Ref.~\cite{Gasenzer:2005ze}). 
As a consequence of the above mentioned effective cut off of the CTP at 
times greater than $\tau$, it will be sufficient, for the time evolution
up to $t=\tau$, to determine the functions $\Gamma^{(n)}_{\tau}$ and thus
the propagator 
\begin{align}
\label{eq:Gtau}
  G_{\tau,ab}
  =i[\Gamma^{(2)}_{\tau}+R_\tau]_{ab}^{-1}.
\end{align}

We restrict ourselves to the case $\phi_a\equiv0$, such that
the action (\ref{eq:Sq}) implies that $\Gamma_{\tau}^{(3)}\equiv0$,
and thus the flow of $\Gamma_\tau^{(1)}$ vanishes.  Moreover, the
equation for the proper two-point function involves, on the right hand
side, only the term containing $\Gamma_\tau^{(4)}$,
\begin{align}
\label{eq:flowGamma2}
  \partial_\tau \Gamma^{(2)}_{\tau,ab}
  &= \frac{i}{2} \int_{{\cal C}}\!
  \Gamma^{(4)}_{\tau,abcd}
  (G_\tau\,[\partial_\tau R_\tau] G_\tau)_{dc}\, ,
\end{align}
see also Fig.~\ref{fig:FlowEqs}.
The term in parentheses stands for the regularised line.
We supplement
Eq.~(\ref{eq:flowGamma2})
with the flow equation for $\Gamma^{(4)}_\tau$, 
which, for $\phi_a\equiv0$, is depicted in Fig.~\ref{fig:FlowEqs}.
This system of equations is still exact. For practical computations
it needs to be closed which can be achieved by truncation or
by supplementing it with equations for one or more higher $n$-vertices
truncated at some higher order.  Here we truncate by neglecting, in
the equation for $\Gamma^{(4)}_\tau$, the term involving
$\Gamma^{(6)}_\tau$,  
\begin{align}
\label{eq:flowGamma4}
\partial_\tau \Gamma^{(4)}_{\tau,abcd} 
  &= -\frac{1}{8}\int_{{\cal C}}\!  \Big\{
  \Gamma^{(4)}_{\tau,abef} G_{\tau,fg}
  \Gamma^{(4)}_{\tau,cdgh}
    \Big\}\,\nonumber\\[2ex]
&\quad \quad\quad \times  (G_\tau [\partial_\tau R_\tau] G_\tau)_{he} \,
  +\, P(a,b,c,d).
\end{align}
$P$ implies a sum corresponding to all permutations of its arguments.
In this way we obtain a closed set of
integro-differential equations for the proper functions up to fourth
order.  As we will show in the following, they allow to derive, for a
particular cutoff time $\tau$, a set of dynamic equations describing
the time evolution of the two- and four-point functions up to time
$t=\tau$.  We emphasise that the only approximation here is the
neglection of the six-point vertex, see Fig.~\ref{fig:FlowEqs}.

\textit{Dynamic equations}.
\label{sec:DynEqs}
For the sharp temporal cutoff $R_\tau$ chosen here the flow equations
can be analytically integrated over $\tau$.
As pointed out above, our cutoff implies the connected two-point function
to vanish at times greater than $\tau$, i.e., it can be written as
\begin{align}
\label{eq:GtauSharp}
  G_{\tau,ab}
  &
  =i\left[\Gamma^{(2)}_\tau\right]^{-1}_{ab}\,
  \theta(\tau-t_a)\,\theta(\tau-t_b)\, ,
\end{align}
where $\theta(\tau)$ evaluates to $0$ for $\tau<0$ and to $1$
elsewhere, and where $t_a$ is the time argument corresponding to the
field index $a$, etc.  Hence, the precise way in which the cutoff
$R_{\tau,ab}$ diverges at $t_a=t_b>\tau$ is chosen such that
$\Gamma^{(2)}_\tau+R_\tau$ is the inverse of $-iG_\tau$ for all times
$t_a,t_b$, see Eq.~(\ref{eq:Gtau}).  Using Eq.~(\ref{eq:GtauSharp})
one finds that
\begin{align}
\label{eq:GRdotG}
  (G_\tau [\partial_\tau R_\tau] G_\tau)_{ab}
  &=-iG_{\tau,ab}\partial_\tau[
  \theta(\tau-t_a)\,\theta(\tau-t_b)].
\end{align}
Note that we have not used the specific choice \eq{eq:Rchoice} for
deriving \eq{eq:GRdotG} but simply the property \eq{eq:GtauSharp},
that is the suppression of any propagation for times $t>\tau$. This
independence of the specific choice of $R_\tau$ leading to
\eq{eq:GtauSharp} is the generic feature of the present approach. 
After inserting Eq.~(\ref{eq:GRdotG}) into Eqs.~(\ref{eq:flowGamma2})
and (\ref{eq:flowGamma4}), we can integrate over $\tau$ and obtain,
after some algebra, the integral equations determining the flow of the
proper functions from $t_0$ to some final time $t$,
\begin{eqnarray}
\label{eq:Gamma2t}
\hspace{-.7cm}\left.\Gamma^{(2)}_{\tau,ab}\right|_{t_0}^t &=&
\frac{1}{2}\int_{t_0,{\cal C}}^t\!
\Gamma^{(4)}_{\tau_{cd},acbd} G_{\tau_{cd},dc},
\\[2ex]
\label{eq:Gamma4t}
\hspace{-.7cm}\left.\Gamma^{(4)}_{\tau,abcd}\right|_{t_0}^t &=&
\frac{i}{2}\int_{t_0,{\cal C}}^t\!
\Gamma^{(4)}_{\tau_{efgh},abef}G_{\tau_{fg},fg}
\nonumber\\[1ex]
&& \hspace{-1cm}\times\
\Gamma^{(4)}_{\tau_{efgh},cdgh}G_{\tau_{eh},he}
  +\, (a\leftrightarrow c) + (a\leftrightarrow d) .
\end{eqnarray}
Double indices imply sums over field components, spatial integrals and
time integrations over the CTP $\cal C$, from $t_0$ to $t$ and back to
$t_0$.  We furthermore introduced 
\begin{align}
\label{eq:tauab}
  \tau_{ab}&=\mathrm{max}\{t_a,t_b\},
  \nonumber\\[2ex]
  \tau_{abcd}
  &=\mathrm{max}\{t_a,t_b,t_c,t_d\}.
\end{align}
The brackets denote terms with the respective indices swapped.

From Eqs.~(\ref{eq:Gamma2t}) and (\ref{eq:Gamma4t}) it is clear that
for the two- and four-point functions to be defined at $\tau=t$ we
need to specify initial functions at $\tau=t_0$.  
We point out that, within the truncation scheme chosen above, we can 
insert any set of proper two- and four-point functions defined in 
their time arguments at and only at $t_0$, as long as 
we set all $n$-vertices for $n=1,3$, and $n>4$ to vanish.  
Our scheme corresponds to a Gaussian initial density matrix $\rho_D(t_0)$ 
since the four-point function has an influence on $\rho_D(t)$ only for $t>t_0$.
Here, we choose the
respective classical proper functions defined by $S$ in
Eq.~(\ref{eq:Sq}).  Hence, the initial two- and four-point functions
entering Eqs.~(\ref{eq:Gamma2t}) and (\ref{eq:Gamma4t}) read
\begin{align}
  \Gamma^{(2)}_{t_0,ab}
  &=S^{(2)}_{ab}=iG^{-1}_{0,ab},
  \\[2ex]
  \Gamma^{(4)}_{t_0,abcd}
  &=S^{(4)}_{abcd}
  =-(2g/{\cal N})(\delta_{ab}\delta_{cd}+\delta_{ac}\delta_{bd}
     +\delta_{ad}\delta_{bc})
  \nonumber\\[1ex]
  &\quad\times\ \delta_{\cal C}(x_a-x_b)\delta_{\cal
  C}(x_b-x_c)\delta_{\cal C}(x_c-x_d).
\end{align}
In order to arrive at a set of dynamic differential equations, we
finally rewrite Eq.~(\ref{eq:Gamma2t}) as
\begin{align}
\label{eq:DynEqG2}
iG_{0,ac}^{-1}G_{\tau_{cb},cb} &= i\delta_{{\cal
    C},ab}-\frac{1}{2}\int_{t_0,{\cal C}}^{\tau_{cb}}\!
\Gamma^{(4)}_{\tau_{de},adce}\,G_{\tau_{de},ed}\,G_{\tau_{cb},cb},
\end{align}
with $\delta_{{\cal C},ab}=\delta_{ab}\delta_{\cal
  C}(x_a-x_b)=\delta_{ab}\delta_{\cal
  C}(t_a-t_b)\delta(\mathbf{x}_a-\mathbf{x}_b)$.
Eq.~(\ref{eq:DynEqG2}) is the dynamic (Dyson-Schwinger) equation for
the connected two-point function $G_{\tau_{ab},ab}$.  
Since $\Gamma^{(2)}_{t_0}=iG_0^{-1}$ represents a differential
operator, the solution of Eq.~(\ref{eq:DynEqG2}) finally requires
another set of boundary conditions to be specified, depending on the
form of the differential operator.  In our case, this is the initial
two-point function, fixed by the one-body density
matrices as well as Bose statistics \cite{Gasenzer:2005ze}.
We finally point out
that we have set $t=\tau_{cb}$ since the flow of $G_{\tau,cb}$ stops
at the maximum of the time arguments $t_b,t_c$.  This can be proven
from the structure of Eqs.~(\ref{eq:Gamma4t}), (\ref{eq:DynEqG2}) but
is more easily seen from the definition (\ref{eq:defZtau}): Once the
hard cutoff $\tau$ has passed the largest time appearing in a
(connected) correlation function, the flow stops since the forward and
backward parts of the CTP over greater times cancel identically in the
functional integral.

Let us assume that $t=t_a$ denotes the present time, at which
Eq.~(\ref{eq:DynEqG2}) determines the further propagation of
$G_{t_a,ab}$ for $t_b\le t_a$ (for $t_b>t_a$ the solution is then
fixed by symmetry).  We point out that all time arguments of the
functions occurring on the right-hand sides of
Eqs.~(\ref{eq:Gamma2t}), (\ref{eq:Gamma4t}), and therefore all cutoff
times $\tau_{ab}$ and $\tau_{efgh}$ are evaluated at times $t'\le t$.
Hence, in accordance with causality, Eqs.~(\ref{eq:Gamma4t}) and
(\ref{eq:DynEqG2}) for a given initial $G_{t_0,ab}(t_0,t_0)$, can be
solved iteratively.
\begin{figure}[tb]
\begin{center}
  \includegraphics[width=0.45\textwidth]{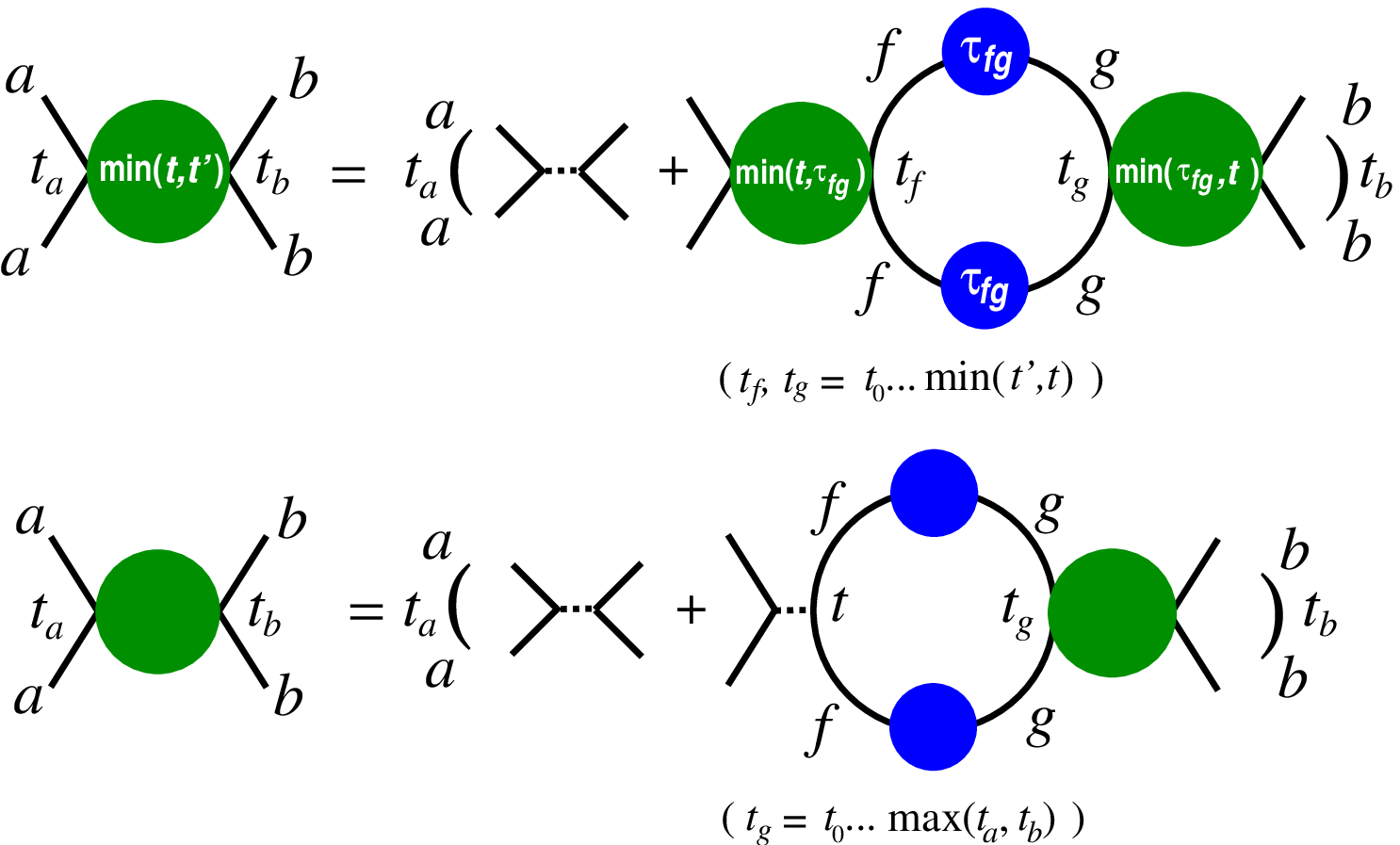}
\end{center}
\vspace*{-3ex} \caption{(color online) The upper equation is the
  $s$-channel projection of Eq.~(\ref{eq:Gamma4t}).  The second
  equation defines the resummed vertex appearing in the
  Dyson-Schwinger equation derived from the NLO $1/{\cal N}$
  approximation of the 2PI effective action
  \protect\cite{Berges2002a}.  The two definitions are identical in
  every order of a perturbative expansion (see text).  Dashed lines
  denote the $s$-channel part of the bare vertex $\Gamma_{t_0,ab}^s$,
  see text.  All other symbols correspond to those in
  Fig.~\ref{fig:FlowEqs}.  Letters on internal lines indicate
  summation over field indices and integration over space and time
  (along the CTP from $t_0$ to $t$ and back).  The integration
  intervals are given in parentheses.  }
\label{fig:Bubblesum}
\end{figure}

This concludes the derivation of our main results, a closed set of
dynamic equations for the two-point correlation function as obtained
from a functional RG approach with a cutoff in real time.  In the
remainder of this article we will concentrate on rederiving the
intensively used dynamic equations obtained from the 2PI effective
action in NLO of an expansion in inverse powers of the number of field
components $\cal N$ and comment on the potential of our approach
beyond this approximation.

\textit{From RG to 2PI next-to-leading order (NLO) $1/{\cal N}$}.
\label{sec:Nonpert}
The mean-field approximation, i.e., the well-known Hartree-Fock
equations are obtained by neglecting the flow of the four-point
according to Eq.~(\ref{eq:Gamma4t}),
$\Gamma^{(4)}_{t}\equiv\Gamma^{(4)}_{t_0}$.  As a consequence, the
flow parameter of $G_{\tau,ab}$ is, for the chosen cutoff, fixed to
$\tau=\tau_{ab}$ and can thus be neglected.
Cf.~Ref.~\cite{Gasenzer:2005ze} for the mean-field equations in the
notation used here.

As a first step beyond mean-field we consider the truncation in which
the $s$-channel scattering diagram is included beyond the mean-field
limit in the loop integral on the right-hand side of
Eq.~(\ref{eq:Gamma4t}).  The result of this section will be that the
obtained equations are equivalent to the 2PI equations in NLO of a
$1/{\cal N}$ expansion \cite{Berges2002a}.  This truncation
corresponds to keeping only one channel of each, the classical vertex
and the one-loop integral term.  The four-vertex then only depends on
two space-time variables and field indices,
$\Gamma^{(4)}_{t,acbd}=\Gamma^{(4)s}_{t,ab}\delta_{ac}\delta_{bd}$,
and enters, in Eq.~(\ref{eq:DynEqG2}), as a cutoff-dependent self
energy
$\Sigma_{\tau_{ab},ab}\equiv\Gamma^{(4)s}_{\tau_{ab},ab}G_{\tau_{ab},ab}$.
We can always write $t=\mathrm{min}(t,t')$, with $t=t'=\tau_{ab}$, for
the cutoff parameter in $\Gamma^{(4)s}_{t,ab}$, see
Fig.~\ref{fig:Bubblesum}, upper line.  Since the parameter in the
(green) vertices on the right hand side is also the minimum of the
maxima of integration times in the adjacent loop and the respective
external time $t$ or $t'$, one can iterate the integral equation in
order to obtain a perturbative series of bubble-chain diagrams
consisting only of classical vertices and full, cutoff-dependent
propagators.  This procedure provides us with a proof that
$\Sigma_{\tau_{ab},ab}$ is \emph{identical} to the NLO 2PI $1/{\cal
  N}$ self energy obtained from the 2PI effective action,
cf.~Ref.~\cite{Berges2002a} and Fig.~\ref{fig:Bubblesum}, lower line.

This proof which exploits the perturbative expansion is lengthy and
will be given elsewhere.  Here we show that the above identity can be
inferred in a comparatively easy way from the topology of the
different terms in the flow equations for the two-, four- and
six-point functions: Consider the untruncated set of equations as
displayed in Fig.~\ref{fig:FlowEqs}.  First, non-$s$-channel
contributions do not generate bubble-chains of the form shown in
Fig.~\ref{fig:Bubblesum}. Second, $\Gamma_\tau^{(6)}$ is one-particle
irreducible. Its contribution to the flow of $\Gamma_\tau^{(4)}$ does
not give rise to bubble-chains, even if inserted recursively into the
first diagram on the right-hand side of the flow equation for
$\Gamma_\tau^{(4)}$. In turn, by dropping the second diagram with the
six-point function and using the $s$-channel truncation, the iterated
flow equation generates only bubble-chain diagrams with full
propagators as lines.  Hence, a $\tau$-integration of this set of flow
equations leads to dynamic equations which include all bubble-chain
contributions and are therefore equal to those obtained to NLO in a
$1/{\cal N}$ expansion of the 2PI effective action \cite{Berges2002a}.
We emphasise that the above topological arguments are generally valid
when comparing resummation schemes inherent in RG equations of the
type of Eq.~\eq{eq:flowGamma}, with those obtained from 2PI effective
actions. This applies, e.g., to equilibrium flows
\cite{Bagnuls:2000ae,Litim:1998nf,Pawlowski:2005xe} and thermal flows
\cite{Litim:1998nf,Blaizot:2006rj}. For a comparison with 2PI results
see Ref.~\cite{Blaizot:2006rj}, for the interrelation of 2PI methods
and RG flows Ref.~\cite{Pawlowski:2005xe}. In this context we would
also like to remark that our approach, by constructions, obeys
Goldstone's theorem as it is solely based on the 1PI effective action
and its derivatives, the proper $n$-point functions. Hence,
spontaneous symmetry breaking always goes together with gapless
excitations in the proper 2-point functions.  For a resolution of this
matter in the 2PI approach see e.g.\
Refs.~\cite{Aarts:2002dj,vanHees:2001ik}.

\textit{Equilibration of a strongly interacting Bose gas}.
\label{sec:Equilibration}
%
\begin{figure}[tb]
\begin{center}
\includegraphics[width=0.23\textwidth]{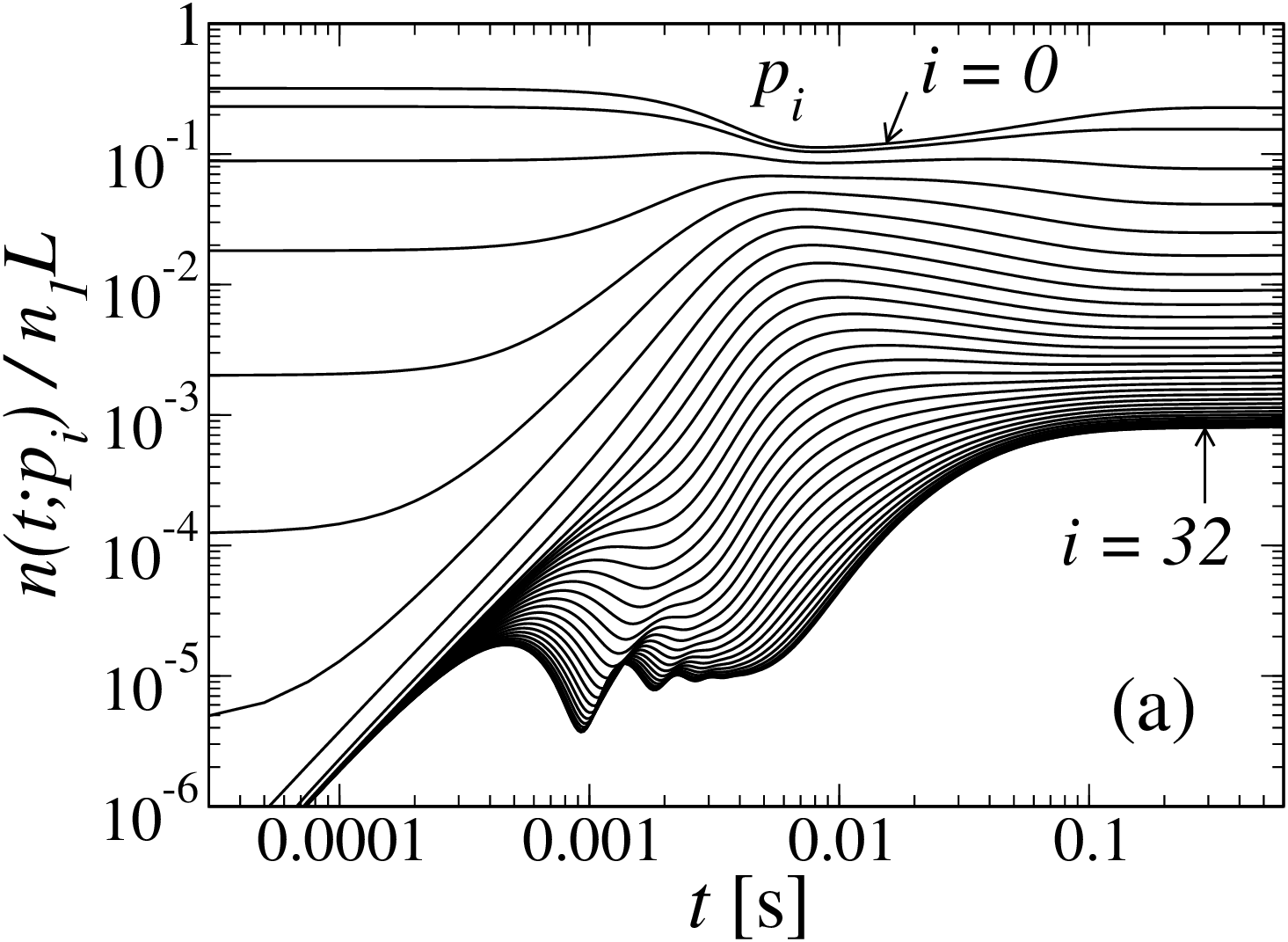}
\includegraphics[width=0.23\textwidth]{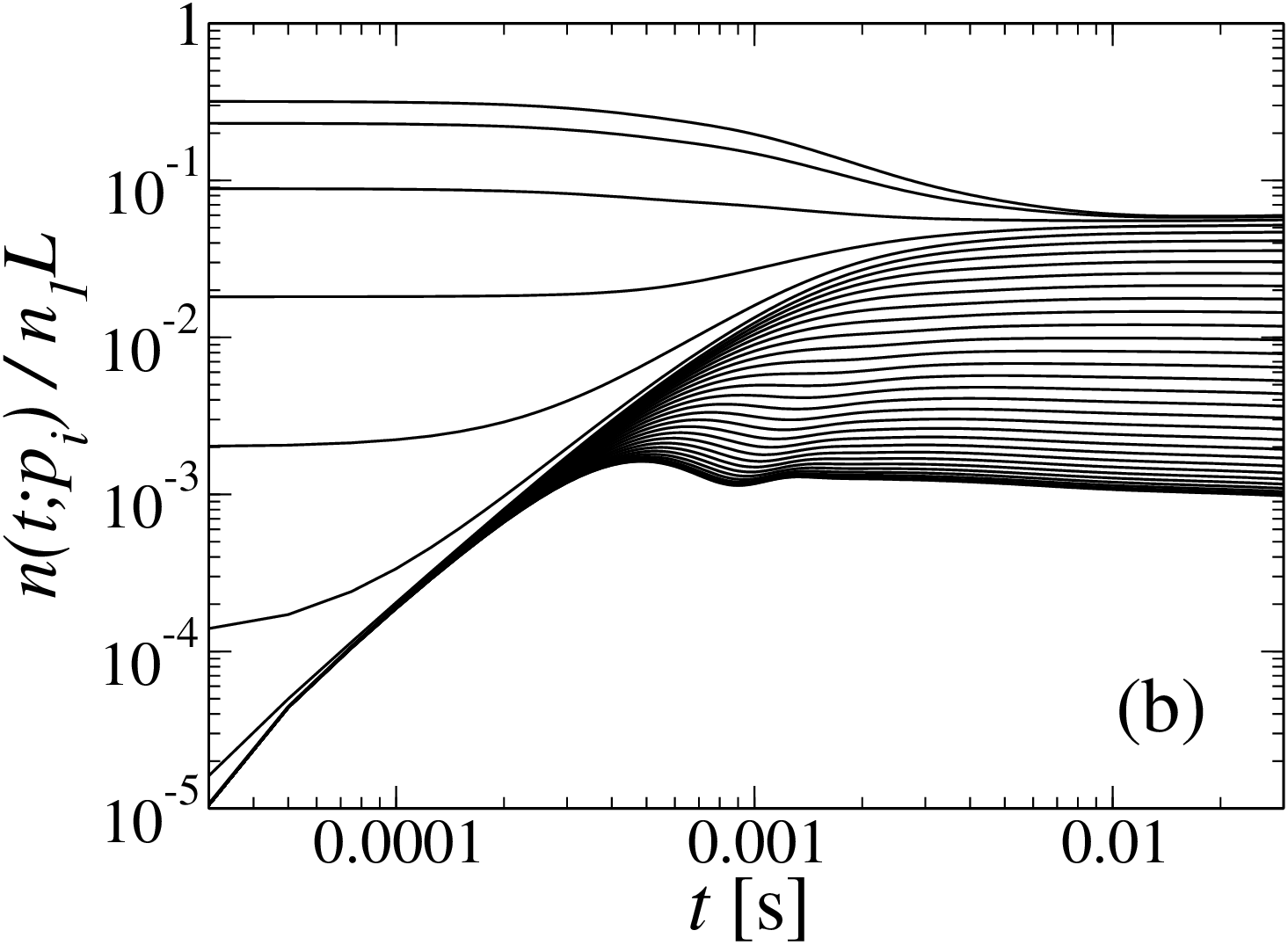}
\end{center}
\vspace*{-3ex} \caption{ Time evolution of momentum-mode occupation
  numbers $n(t;p)/n_1L$ of a uniform 1D Bose gas, normalised by the
  total number $n_1L$ of atoms in the box.  The initial
  far-from-equilibrium state corresponds to a Gaussian momentum
  distribution. Panel (a) describes the evolution of a weakly
  interacting gas, $\gamma=1.5\cdot10^{-3}$, panel (b) that of a
  strongly interacting gas with $\gamma=15$.  }
\label{fig:timeevolution}
\end{figure}
We close our paper with an illustration of the non-perturbative nature
of the dynamic equations obtained in the $s$-channel approximation,
which are, as shown, equivalent to the 2PI equations in NLO 1/${\cal
  N}$ approximation.  For this we face the long-time evolution of a
weakly interacting non-relativistic Bose gas obtained in
Ref.~\cite{Berges2007a}, for which $1/{\cal N}=1/2$, with that of a
strongly interacting one.  These results illustrate, what is supported
by the non-perturbative approach presented here as well as by
benchmark tests for special-case systems \cite{Aarts2002a}, viz.~that
the $s$-channel or 2PI $1/{\cal N}$ approximation is applicable for
strong coupling also for small ${\cal N}$ where an expansion in powers
of $1/{\cal N}$ would seem questionable.  The uniform gas in one
spatial dimension starts from a non-equilibrium state which is assumed
to be fully described by a non-vanishing two-point function
$G_{aa}(t_0,t_0;p)=n(t_0;p)+1/2$, $a=1,2$. The initial momentum
distribution of atoms in the gas is
$n(t_0;p)=({n_1}{\sqrt{\pi}\sigma})\exp({-p^2/\sigma^2})$, with
$\sigma=1.3\cdot10^5\,$m$^{-1}$.  The computations were done as
described in Ref.~\cite{Berges2007a}, on a grid with $N_s=64$ points
spaced by $a_s=1.33\mu$m.  Fig.~\ref{fig:timeevolution} shows the
evolution for a (a) weakly interacting sodium gas, with line density
$n_1=10^7$ atoms$/$m and interaction strength $g=\hbar^2\gamma n_1/m$,
$\gamma=1.5\times10^{-3}$ and (b) a strongly interacting one with
$\gamma=15$, $n_1=10^5/$m. The final momentum distributions closely
resemble Bose-Einstein, i.e., thermal distributions
$n(t;p)\propto1/[\exp(\beta[\omega_p-\mu])-1]$, with the dispersion
$\omega_p$ extracted from the numerical data \cite{Berges2007a}.

\textit{Conclusions}.
\label{sec:concl}
Using functional RG techniques we have derived a theory of
far-from-equilibrium quantum field dynamics. The time evolution of the
system is generated through the flow with a cutoff time introduced in
the generating functional for correlation functions.  In a truncation
of the flow equations we recover the dynamic equations as known from
the $1/\cal N$-expansion of the 2PI effective action.  These results
lead us to the following conclusions: Although sub-leading terms in
the expansion of the 2PI effective action are suppressed with
additional powers of $1/\cal N$ they contain bare couplings, and,
despite their resummation, the approximation seems formally
questionable for strong couplings if $g/\cal N$ is large.  Presently,
an intense discussion focuses on the question to what extent the NLO
$1/{\cal N}$ approximation is applicable for large interaction
strengths, drawing from benchmark tests for special-case systems
\cite{Aarts2002a}.  Our results provide analytical means to find the
conditions for the validity of this approximation since no
perturbative ordering in bare couplings is involved, and the remaining
truncation in the Green functions can be tested for self-consistency.

T.G. thanks J.~Berges for collaboration on related work and the
Deutsche For\-schungs\-gemeinschaft for support. We thank S. Ke\ss ler
for discussions, and W.~Wetzel for his continuing help with the
computational resources used in this work.
\\[-6.3ex]

\bibliography{mybib,additions}

\end{document}